# Flavor Mixing

## of

## Quarks and Neutrinos

**H. Fritzsch**

**Department of Physics**

**University of Munich**

**Munich, Germany**

The phenomena of flavor mixing of the quarks and leptons depend on 22 parameters: the three masses of the charged leptons, the three neutrino masses, the six quark masses, the four parameters in the mixing matrix of the quarks, the four parameters in the mixing matrix of the leptons and two phase parameters (in case, the neutrino masses are Majorana masses).

I shall describe the flavor mixing of the quarks and leptons using "texture zero mass matrices" (ref.1, 2). In this case the flavor mixing angles are determined by the ratios of the masses of the quarks and leptons.

Mass matrices with "texture zeros" cannot be obtained in the Standard Model, but in extended models, if discrete or continuous symmetries are present. For example, such symmetries are present in left-right symmetric electroweak theories. If the grand unification is described by the gauge group SO(10), the electroweak sector is given by a left-right symmetric theory.

First I shall discuss the case of two generations of quarks. In this case the flavor mixing is described by one angle, the Cabibbo angle. The "texture zero mass matrices" for two generations have the following structure:

$$\begin{pmatrix} 0 & a \\ a^* & b \end{pmatrix} \qquad (1)$$

If these matrices are diagonalized, one obtains the mass eigenvalues:

$$\begin{pmatrix} 0 & a \\ a^* & b \end{pmatrix} \succ \begin{pmatrix} -m_1 & 0 \\ 0 & m_2 \end{pmatrix} \qquad (2)$$

The associated rotation angle is given by the ratio of the mass eigenvalues:

$$\theta \approx \sqrt{\frac{m_1}{m_2}} \qquad (3)$$

The rotation angles for the two mass matrices are given by the quark mass ratios:

$$\theta_d \approx \sqrt{\frac{m_d}{m_s}} \qquad \theta_u \approx \sqrt{\frac{m_u}{m_c}} \qquad (4)$$

The Cabibbo angle is given by the difference of the two rotation angles and an unknown phase parameter:

$$\theta_c \cong \left| \theta_d - e^{i\phi} \theta_u \right| \qquad (5)$$

We shall use the following quark masses, normalized at 1 GeV:

$$m_d \approx 7.3 \quad MeV$$
$$m_s \approx 150 \quad MeV$$
$$m_u \approx 5.2 \quad MeV$$
$$m_c \approx 1050 \quad MeV$$

$$\sqrt{\frac{m_d}{m_s}} \approx 0.22$$

$$\sqrt{\frac{m_u}{m_c}} \approx 0.07 \qquad (6)$$

The Cabibbo angle depends on the phase parameter. The observed value of the Cabibbo angle fixes the phase parameter to 90 degrees.

We consider now the case of three generations. The flavor mixing is described by three angles and one phase parameter (ref. 3). The phase parameter describes the CP-violation. We shall use a specific mixing matrix, introduced by Xing and the author (ref. 4):

$$V = \begin{bmatrix} c_u & s_u & 0 \\ -s_u & c_u & 0 \\ 0 & 0 & 1 \end{bmatrix} \cdot \begin{bmatrix} e^{-i\phi} & 0 & 0 \\ 0 & c & s \\ 0 & -s & c \end{bmatrix} \cdot \begin{bmatrix} c_d & -s_d & 0 \\ s_d & c_d & 0 \\ 0 & 0 & 1 \end{bmatrix} \quad (7)$$

The angle with the index u describes the mixing in the u-c sector, the angle with the index d the mixing in the d-s sector. The angle with no index describes the mixing between the heavy quarks.

For three generations the mass matrices with texture zeros are as follows:

$$M = \begin{bmatrix} 0 & A & 0 \\ A^* & B & C \\ 0 & C^* & D \end{bmatrix} \quad (8)$$

Two of the three mixing angles are determined by the quark mass ratios:

$$\tan \theta_d = \sqrt{\frac{m_d}{m_s}}$$

$$\tan \theta_u = \sqrt{\frac{m_u}{m_c}} \quad (9)$$

Below we give the theoretical and experimental values for these two angles - the agreement between the theoretical and experimental values is remarkable:

$$\theta_d \approx 13.0 \pm 0.4^o \qquad \theta_u \approx 5.0^o \pm 0.7^o$$
$$Exp: 11.7^o \pm 2.6^o \qquad Exp: 5.4^o \pm 1.1^o \tag{10}$$

The CKM matrix is a unitary matrix, thus the scalar product of the first column and the third column must be zero. This leads to a triangular relation in the complex plane (unitarity triangle). With the "texture zero mass matrices" all angles of the unitaritiy triangle can be calculated. In our approach the unitarity triangle is the same triangle, which we discussed in the case of two generations. The Cabibbo angle was the difference of the two angles above and a phase parameter, which had to be 90 degrees. This phase parameter is the angle alpha in the unitarity triangle. The experimental value of this angle is consistent with 90 degrees.

We can also calculate the angle beta:

$$\tan \beta = \frac{\sin \theta_u \cos \theta_d}{\cos \theta_u \sin \theta_d}$$

==>

$$\sin 2\beta \cong 0.663 \tag{11}$$

$$Exp: \sin 2\beta = 0.681 \pm 0.025$$

The theoretical value agrees very well with the value, observed in the experiments.

In the following we shall discuss the flavor mixing of the leptons, which mainly manifests itself in neutrino oscillations. The flavor mixing of leptons is described by a mixing matrix V analogous to the CKM matrix:

$$V = \begin{pmatrix} V_{1e} & V_{2e} & V_{3e} \\ V_{1\mu} & V_{2\mu} & V_{3\mu} \\ V_{1\tau} & V_{2\tau} & V_{3\tau} \end{pmatrix} \tag{12}$$

In the case of the leptons the matrix V is a product of a phase matrix P and of a unitary matrix U:

$$V = U \cdot P$$

$$P = \begin{pmatrix} e^{i\rho} & 0 & 0 \\ 0 & e^{i\sigma} & 0 \\ 0 & 0 & 1 \end{pmatrix}$$

(13)

$$U = \begin{bmatrix} \cos\theta_l & \sin\theta_l & 0 \\ -\sin\theta_l & \cos\theta_l & 0 \\ 0 & 0 & 1 \end{bmatrix} \cdot \begin{bmatrix} e^{-i\varphi} & 0 & 0 \\ 0 & \cos\theta & \sin\theta \\ 0 & -\sin\theta & \cos\theta \end{bmatrix} \cdot \begin{bmatrix} \cos\theta_v & -\sin\theta_v & 0 \\ \sin\theta_v & \cos\theta_v & 0 \\ 0 & 0 & 1 \end{bmatrix}$$

The three angles in the matrix U are the atmospheric angle, the solar angle and the reactor angle:

$$\theta \approx \theta_{at} \qquad \theta_v \approx \theta_{sun} \qquad \theta_l \approx reactor - angle$$

The oscillation experiments in Kamioka and at the Sudbury Neutrino Observatory provide information about the mixing angles and the mass differences of the neutrinos:

$$31.7^o \leq \theta_{sun} \leq 36.3^o$$

$$38^o \leq \theta_{at} \leq 52^o$$

$$\Delta m_{21}^2 \approx 7.6 \cdot 10^{-5} eV^2 \qquad (14)$$

$$\Delta m_{32}^2 \approx 2.4 \cdot 10^{-3} eV^2$$

We assume that the lepton masses are also given by the "texture zero mass matrices". One finds the following relations between the angles and the masses:

$$\tan 2\theta_l = \frac{2\sqrt{m_e m_\mu}}{m_\mu - m_e} \cong 0.0695$$

$$\tan 2\theta_\nu = \frac{2\sqrt{m_1 m_2}}{m_2 - m_1} \qquad (15)$$

The electron and muon masses determine the reactor mixing angle. The solar angle is determined by the masses of the first and second neutrino. Since the solar angle is measured, we can deduce the ratio of the masses of the first and second neutrino:

$$m_1 / m_2 \approx 0.42^{+0.12}_{-0.04} \qquad (16)$$

This relation and the experimental results for the mass differences determine the neutrino masses. We obtain in units of electron volt:

$$m_1 = 0.004 \pm 0.0012 \quad (17)$$
$$m_2 = 0.011 \pm 0.002$$
$$m_3 = 0.051 \pm 0.007$$

We find a normal hierarchy for the neutrino masses, not an inverted hierarchy, but the mass hierarchy is relatively weak, compared to the mass hierarchy for the charged leptons and for the quarks.

We can also calculate the matrix element of V, which is relevant for the reactor angle:

$$V_{e3} = \sin\theta_{13} = \sin\theta \sin\theta_l \cong 0.707 \sqrt{\frac{m_e}{m_\mu}} = 0.049 \pm 0.006$$

$$\sin^2 2\theta_{13} = 0.0096 \pm 0.0022 \quad (18)$$

The limit of the CHOOZ experiment for this matrix element is 0.2 – we expect 0.05. The DAYA BAY experiment is expected to decrease the limit to 0.045. If our ansatz of the "texture zero" mass matrices is correct, the DAYA BAY experiment will observe an effect.

If the neutrinos have Majorana masses, one might observe the neutrinoless double beta decay. The limit on the effective neutrino mass from the Cuoricino experiment in the Gran Sasso Laboratory is about 0.23 eV. We can calculate the effective neutrino mass. It receives contributions from all three neutrinos, but the sum is only 0.012 eV. Thus one would have to improve the limit by a factor 20 in order to see an effect.

**We conclude:** The "texture zero mass matrices" for the quarks and leptons describe very well the flavor mixing of the quarks and leptons. We can calculate the angles of the unitarity triangle. We expect the angle alpha of the unitarity triangle to be 90 degrees. The masses of the neutrinos can be calculated – they are very small, the largest neutrino mass is 0.05 eV. We calculated the matrix element of the mixing matrix, relevant for the reactor mixing angle. It can be measured in the near future in the DAYA BAY experiment.